\documentstyle[eqsecnum,aps,amssymb]{revtex}

\def\a{\alpha}
\def\b{\beta}

\def\e{\varepsilon}
\def\th{\vartheta}
\def\l{\lambda}
\def\r{\rho}

\def\s{\sigma}

\def\ph{\varphi}

\def\square{{\vcenter{\vbox{\hrule height.4pt \hbox{\vrule width.4pt 
height1.45ex \kern1.45ex \vrule width.4pt}
\hrule height.4pt}}}}

\def\qed{\penalty10000\hfill$\square$\par\goodbreak\medskip}

\def\proof{\medskip\noindent{\it Proof.} }

\def\definition{\bigbreak\noindent{\bf Definition.} }
 
\def\B{{\cal B}}
\def\H{{\cal H}}
\def\K{{\cal K}}
\def\L{{\cal L}}
\def\S{{\cal T}}

\def\NN{{\Bbb N}}
\def\RR{{\Bbb R}}

\def\CC{{\Bbb C}}

\def\tr{{\rm tr}\,}

\def\Col{|\alpha_0|^2 \ketbra{\psi_0} + |\alpha_1|^2 \ketbra{\psi_1}}

\def\tuple#1_#2{#1_1,#1_2,\ldots,#1_{#2}}
\def\tup#1_#2{#1_1,\ldots,#1_{#2}}
\def\bra#1{\langle#1|}
\def\ket#1{|#1\rangle}
\def\ketbra#1{|#1\rangle\langle#1|}

\def\inp#1#2{\langle#1,#2\rangle}
\def\norm#1{\|#1\|}
\def\norms#1{\|#1\|^2}

\def\abss#1{|#1|^2}
\def\one{{\bf 1}}
\def\ten{\otimes}

\newtheorem{theorem}{Theorem}
\newtheorem{lemma}[theorem]{Lemma}
\newtheorem{corollary}[theorem]{Corollary}
\newtheorem{proposition}[theorem]{Proposition}

\begin{document}
\draft

\title{Information Transfer Implies State Collapse}

\author{B.~Janssens and H.~Maassen}
\address{Mathematisch Instituut, 
         Radboud Universiteit Nijmegen, 
         Toernooiveld 1, 
         6525 ED Nijmegen, 
         The Netherlands, 
         {\tt maassen@math.kun.nl, basjanss@sci.kun.nl}}

\date{February 16, 2006}
\maketitle
\begin{abstract}\noindent We attempt to clarify certain puzzles concerning 
state collapse and decoherence.
In open quantum systems decoherence is shown to be a necessary
consequence of the transfer of information to the outside;
we prove an upper bound for the amount of coherence which can survive
such a transfer.
We claim that in large closed systems decoherence has never been observed,
but we will show that it is usually harmless to assume its occurrence.
An independent postulate of state collapse over and above
Schr\"odinger's equation and the probability interpretation
of quantum states, is shown to be redundant.
\end{abstract}

\section{Introduction}\noindent
In its most basic formulation,
quantum theory encodes the preparation of a system in a pure quantum state,
a unit vector $\psi$ in a Hilbert space $\H$. 
Observables are modelled by (say, nondegenerate) self-adjoint operators
on $\H$.
The expectation value of an observable $A$ in a state $\psi$
is given by $\inp{\psi}{A \psi}$. 
If $a$ is an eigenvalue of $A$ and $\psi_a$ a unit eigenvector,
and information concerning $A$ is somehow extracted from the system,
then the probability for the value $a$ to be observed
is $|\inp{\psi_a}{\psi}|^2$.
If this observation is indeed made,
then the subsequent behaviour of the system is predicted using
the pure state $\psi_a$.
This is called \emph{state collapse}.
It follows that,
if the information extraction has taken place 
but the information on the value of $A$ is disregarded,
then the subsequent behaviour can be described optimally using a mixture
of eigenstates.
This is called \emph{decoherence}. 
In this paper we substantiate the following claim concerning
decoherence and state collapse.
\begin{center}
{\sl
Decoherence is only observed in open systems, where it is a necessary\\
consequence of the transfer of information to the outside.\\}
\end{center}

\noindent
So the observed occurrence of decoherence does not contradict the unitary
time evolution postulated by quantum mechanics,
since open systems do not evolve unitarily.
Decoherence can be explained in quantum theory
by embedding the quantum system into a larger, closed whole,
which in itself evolves unitarily.
This is well-known (see e.g. \cite{Neu}).
We add the observation that
decoherence is not only a {\em possibility} for an open system,
but a {\em necessary consequence}
of the leakage of information out of the system.
We prove an inequality relating the decoherence between two pure states
to the degree in which a decision between the two is possible
by a measurement outside.
This is the content of Theorem \ref{cor5} in section \ref{open}.

\noindent
Also,
we have claimed that one {\it does not} actually observe decoherence
in closed macroscopic systems.
First of all, 
most of the systems that are ever observed are actually open,
since it is extremely difficult to shield large systems from
interaction.
But more to the point,
the difference between coherence and decoherence can only 
be seen by measuring some highly exotic `stray observables'
which are almost always forbiddingly hard to observe.
And indeed,
in those rare cases where experimenters have succeeded in measuring them,
ordinary unitary evolution was found, not decoherence. 
(See \cite{Arn}, \cite{Fri}, \cite{Wal}.)

\noindent
We illustrate the latter point in section \ref{oester},
where we show that the measurement of two classes of observables 
\emph{can not} reveal the difference between coherence and decoherence:
a class of microscopic observables and a large class of macroscopic
observables.
Take as an example a volume of gas.
Microscopic observables such as the position of one particular atom in a gas, 
only relate to a small fraction of the system.
Macroscopic observables like the center of mass of the gas,
are the average over a large number of microscopic
observables.
Belonging neither to the macroscopic nor to the microscopic class, 
the `stray observables'
referred to above describe detailed correlations between 
large numbers of atoms in the gas. 
This kind of information is  experimentally almost inaccessible.

\noindent
Driving home our point concerning decoherence in closed systems:
coherent superpositions of macroscopically distinguishable states
are not the strange monsters produced by a quantum theory applied
outside its domain.
They are, on the contrary, everyday occurrences which,
however, {\it can not} be distinguished
from the more classical incoherent superpositions in practice,
and can therefore always be regarded as such.


\section{Abstract Information Extraction}\label{perse}\noindent
Quantum phenomena are inherently stochastic.
This means that, if quantum systems are prepared in identical ways,
then nevertheless different events may be observed.
A quantum state describes an ensemble of
physical systems, e.g. a beam of particles,
and is modelled by a normalized 
trace-class operator $\rho$ on the Hilbert space.
The expectation value of an observable $A$ in the state $\rho$ 
is then $\tr(\rho A)$.

An information extraction or measurement on a quantum state is to be
considered as the partition of such an ensemble into subensembles,
each subensemble corresponding to a measurement outcome.
Let us, in the present section,
not wonder {\it how} the splitting of ensembles can be described 
by quantum theory,
but let us see what such an information extraction, if it can be done,
will entail for the subsequent behaviour of the subensembles.
Note that this process may serve as part of the preparation
for further experiments on the system, so that it must again lead to a state.

\subsection{Information Extraction}\label{secIESC}\noindent
For simplicity let us assume that only two outcomes can occur, labelled 
0 and 1, say with probabilities $p_0$ and $p_1$.
The ensemble is then split in two parts,
described by their respective states $\rho_0$ and $\rho_1$.
The map
\begin{equation}
M:\rho\mapsto  p_0 \rho_0\oplus p_1 \rho_1$$
\label{Measure}
\end{equation}
must be normalized, affine and positive.
Indeed, normalization is the property 
that $p_0 + p_1 = 1$, and 
positivity is the requirement that states must be mapped to states.
The affine property entails that for all states $\rho$ and $\theta$
on the original system, and for all $\lambda\in[0,1]$,
   $$M(\l\rho+(1-\l)\th)=\l M(\rho)+(1-\l)M(\th)\;.$$
This follows from the physical principle that a system which is prepared
in the state $\r$ with probability $\l$ and in the state $\th$ with probability
$1-\l$, say by tossing a coin,
can not be distinguished from a physical system in the state
$\l\rho+(1-\l)\th$.
We emphasize that indeed this is a {\it physical} principle,
not a matter of definitions.
It states, for instance, that a bundle of particles
having 50\% spin up and 50\% spin down can not be distinguished from
a bundle having 50\% spin left and 50\% spin right.
This is a falsifyable statement.

\subsection{State Collapse}\noindent
The above elementary observations are sufficient to prove that information
extraction implies state collapse.
If $M$ distinguishes perfectly between
the pure states $\psi_0$ and $\psi_1$, then of course
$p_0=1$ in case $\rho=\ketbra{\psi_0}$, and
$p_1=1$ if $\rho=\ketbra{\psi_1}$.
\begin{proposition}\label{abscol}
Let $\S(\H)$ denote the space of trace class operators on a Hilbert space
$\H$,
and let the map
$M:\S(\H)\to\S(\H)\oplus\S(\H):\rho\mapsto M_0(\rho)\oplus M_1(\rho)$
be the linear extension of some normalized, affine and positive map
on the states.
Suppose that unit vectors $\psi_0$ and $\psi_1$ exist such that
\begin{equation}
     M\bigl(\ketbra{\psi_0}\bigr)=M_0\bigl(\ketbra{\psi_0}\bigr)\oplus0
      \qquad\hbox{and}\qquad
     M\bigl(\ketbra{\psi_1}\bigr)=0\oplus M_1\bigl(\ketbra{\psi_1}\bigr)\;.
\label{fullsplit}
\end{equation}
Then we have $M\bigl( \ket{\psi_0} \bra{\psi_1} \bigr)
= M\bigl( \ket{\psi_1} \bra{\psi_0} \bigr) = 0$.
\end{proposition}
\proof 
The positivity of $M$ yields 
$M\bigl( \ketbra{\e e^{i\ph} \psi_0 + \psi_1} \bigr) \geq 0$
as an operator inequality.
In particular, the 0-th component must be positive.
As $M_0 (\ketbra{\psi_1}) =0$, it follows that for all
$\e, \phi \in \RR$, we have
$\e^2 M_0 \bigl( \ketbra{\psi_0}\bigr) + 
\e \left( e^{i\ph} M_0 \bigl( \ket{\psi_0}\bra{\psi_1}\bigr) + 
e^{-i\ph} M_0 \bigl( \ket{\psi_1}\bra{\psi_0}\bigr) \right) 
\geq 0$.
Taking the limit $\e \downarrow 0$ yields  
$\left( e^{i\ph} M_0 \bigl( \ket{\psi_0}\bra{\psi_1}\bigr) + 
e^{-i\ph} M_0 \bigl( \ket{\psi_1}\bra{\psi_0}\bigr) \right) 
\geq 0$ for all $\ph \in \RR$. In particular for 
$\ph = 0,\frac{\pi}{2},\pi,\frac{3\pi}{2}$, implying
$M_0 \bigl( \ket{\psi_0}\bra{\psi_1} \bigr) = 
M_0 \bigl( \ket{\psi_1}\bra{\psi_0} \bigr) = 0$.

Exchanging the roles of $\psi_0$ and $\psi_1$ in the argument above
results in 
$M_1 \bigl( \ket{\psi_0}\bra{\psi_1} \bigr) = 
M_1 \bigl( \ket{\psi_1}\bra{\psi_0} \bigr) = 0$, proving the proposition.
\qed
\noindent We may draw two conclusions from Proposition \ref{abscol}.
The first is that, 
for all $\ket{\psi}=\a_0\ket{\psi_0}+\a_1\ket{\psi_1}$, we have 
\begin{equation}\label{collapse}
(M_0 + M_1)\bigl(\ketbra{\psi}\bigr) = (M_0 + M_1)\bigl(\Col \bigr)\,.
\end{equation}
In words: for the prediction of events {\it after} the splitting
of the ensemble in two,
it no longer matters whether {\it before} the splitting
the system was in the pure state
$\ketbra {\a_0 \psi_0 + \a_1 \psi_1}$ 
or in the mixed state 
$\abss{\a_0} \ketbra{\psi_0} + \abss{\a_1} \ketbra{\psi_1}$.
This phenomenon, which is a direct consequence of the structure
(\ref{Measure}) of the measurement process,
we will call {\it decoherence}.
   
The second conclusion from Proposition \ref{abscol} is the following.
For all $\ket{\psi}=\a_0\ket{\psi_0}+\a_1\ket{\psi_1}$, we have 
   \begin{equation}M \bigl(\ketbra\psi\bigr)=
        \abss{\a_0} M_0 \bigl(\ketbra{\psi_0} \bigr) \oplus 
		\abss{\a_1} M_1 \bigl(\ketbra{\psi_1}\bigr)\;.
   \end{equation}
In words: if an ensemble is split in two parts, then the `0-ensemble'
will further behave as if the system had been in state $\psi_0$ instead of
$\psi$ prior to splitting, and the 
`1-ensemble' as if it had been in state $\psi_1$ instead of $\psi$.   
This phenomenon will be called {\it collapse}.

Throughout this article, we will maintain a
sharp distinction between the collapse $M : \S(\H) \to \S(\H) \oplus \S(\H)$ 
and the decoherence $(M_0 + M_1) : \S(\H) \to \S(\H) $.
The former represents the splitting of an ensemble in two parts by means of
measurement, whereas the latter represents the splitting and subsequent 
recombination of this ensemble.  

\section{Open Systems} \label{open}\noindent
A decoherence-mapping $(M_0 + M_1):\S(\H)\to\S(\H)$
maps the pure state $\ketbra{\a_0 \psi_0 + \a_1 \psi_1}$
and the mixed state 
$|\a_0|^2\ketbra{\psi_0} + |\a_1|^2\ketbra{\psi_1}$ to the same final state.
Since unitary maps preserve purity, 
there can not exist 
a unitary map $U:\H\to \H$ such
that for all $\rho\in\S(\H)$:
   $$(M_0 + M_1)(\rho)=U\rho U^*\;.$$
However, according to Schr\"odinger's equation the development 
of a closed quantum system is given by a unitary operator.
We conclude that the 
decoherence (\ref{collapse}) 
is impossible in a closed system.
On the other hand decoherence is
a well known and experimentally confirmed phenomenon. 

We will therefore consider open systems,
i.e. quantum systems which do not obey the Schr\"odinger equation, but are part of a
larger system which does. 
It has often been pointed out (e.g. \cite{Neu}, \cite{Zur})
that decoherence can well occur in this situation,
provided that states are only evaluated on the observables of the smaller system.
We are more ambitious here: 
we shall prove that this form of `local' decoherence is not just a
\emph{possible}, but an an \emph{unavoidable} consequence
of information-transfer out of the open system.


\subsection{Unitary Information Transfer and Decoherence}\label{ITSC}\noindent
We assume that the open system has Hilbert space $\H$,
and that its algebra of observables is given by $B(\H)$,
the bounded operators on $\H$.
We may then assume that the larger system has Hilbert space $\K\ten\H$,
since the only way to represent $\B(\H)$ on a Hilbert space
is in the form $A\mapsto\one\ten A$ \cite{Takesaki}.
We may think\footnote{Sometimes it may happen,
as for instance in fermionic systems,
that the observables of the ancilla do not all commute
with those of the open system.
Also the observable algebra on $\K$ may be smaller than
$\B(\K)$,
but we will neglect these complications here.} of 
$\B(\K)$ as the observable algebra of some ancillary system 
in contact with our open quantum system. 
In this context, $\H$ will be referred to
as the `open system', $\K$ as the `ancilla' and $\K \ten \H$
as the `closed system'.

We couple the system to the ancilla during a finite time interval $[\, 0, t \,]$.
Let $\tau \in \S(\K)$ denote the state of the ancilla at time 0,
and $\rho  \in \S(\H)$ that of the small system. 
The effect of the interaction is described by a unitary operator
$U:\K\ten\H\to\K\ten\H$,
and the state of the pair at time $t$ is given by 
$U(\tau \ten \rho)U^* \in \S(\K \ten \H)$.
For convenience, we will define the information transfer map $T : \S(\H) \to \S(\K\ten\H)$ by
$ T(\rho) := U(\tau\ten\rho) U^*\,.$

\subsubsection{Decoherence}\label{zwaan}\noindent
In the above setup, we are interested in distinguishing whether the open system $\H$
was in state $\ket{\psi_0}$ or $\ket{\psi_1}$ at time 0.
This can be done if
there exists a `pointer observable' $B\ten \one$ in the ancilla $B(\K)$
which takes average value $b_0$ in state $T(\ketbra{\psi_0})$ 
and $b_1$ in state $T(\ketbra{\psi_1})$. 
By looking only at the ancilla $\K$ at time $t$, we are then able to 
gain information on the state of the open system $\H$ at time 0. 
We say that information is \emph{transferred} from $\H$ to $\K$.

Under these circumstances, we wish to prove that 
decoherence occurs on the open
system.  
We prepare the ground by proving the following lemma.

\begin{lemma}\label{lem4}.
Let $\th_0,\th_1$ be unit vectors in a Hilbert space $\L$,
and let $A$ and $B$ be bounded self-adjoint operators on $\L$ satisfying
$\norm{[A,B]}\le\delta\norm{A}\cdot\norm{B}.$
For $j=0$ or $1$,
let $b_j := \inp{\theta_j}{B\theta_j}$ denote the expectation and 
$\s_j^2 := \inp{\theta_j}{B^2 \theta_j} - \inp{\theta_j}{B\theta_j}^2$ 
the variance of $B$ in the state $\th_j$.
Then, if $b_0\ne b_1$,
 $$\bigl|\inp{\th_0}{A\th_1}\bigr|
   \le{{\delta\norm B+\s_0+\s_1}\over{|b_0-b_1|}}\norm A\;.$$
\end{lemma}
\proof 
Since
 $\norms{(B-b_j)\th_j}=\inp{\th_j}{(B-b_j)^2\th_j}=\s_j^2,$
we have, by the Cauchy-Schwarz inequality,
   $$\bigl|(b_0-b_1)\inp{\th_0}{A\th_1}\bigr|
     =\bigl|\inp{\th_0}{\bigl(A(B-b_1)-(B-b_0)A+[B,A]\bigr)\th_1}\bigr|
     \le\norm A(\s_1+\s_0)+\delta\norm A\cdot\norm B\;.$$
\qed
\noindent Note that, for $\delta = \sigma_0 = \sigma_1 = 0$, Lemma \ref{lem4}
merely states that commuting operators respect each other's eigenspaces.
We proceed to prove that information transfer causes 
decoherence on the open system. (See \cite{scriptie}.)
\begin{theorem}\label{cor5}
Let $\psi_0$ and $\psi_1$ be mutually orthogonal unit vectors in a Hilbert
space $\H$,
and let $\tau \in \S(\K)$ be a state on a Hilbert space $\K$.
Let $U : \K \ten \H \rightarrow \K \ten \H$ be unitary and define
$T : \S(\H) \rightarrow \S(\K \ten \H)$ by
$T(\rho) = U (\tau \ten \rho) U^* $.
Let $B$ be a bounded self-adjoint operator on $\K\ten\H$, 
and denote by $b_j$ 
and $\s_{j}^{2}$  its expected value  and variance 
in the state $T(\ketbra{\psi_j})$ for $j = 0,1$.
Suppose that $b_0 \neq b_1$. 
Then for all $\psi=\a_0\psi_0+\a_1\psi_1$ with $|\a_{0}|^2 + |\a_{1}|^2 = 1$
and for all bounded self-adjoint operators $A$ on $\K \ten \H$ such that
$\norm{[A,B]}\le\delta\norm{A}\cdot\norm{B}$, we have
\begin{equation} \label{margel}
   \Bigl| \tr \bigl( T(\ketbra{\psi}) A \bigr) -
   \tr \Bigl( T\bigl( \Col ) \bigr) A \Bigr) \Bigr| 
   \le
   {{\delta\norm B+\s_0+\s_1}\over{|b_0-b_1|}}\norm A \,.
\end{equation}
\end{theorem}
\proof 
First, we prove \ref{margel} in the special case that $\tau = \ketbra{\ph}$ 
for some vector $\ph \in \K$. 
We introduce the notation $\theta_j := U(\ph\ten\psi_j)$. 
Recall that the expectation of $B$ is given by
$ b_j = \tr \bigl( T(\ketbra{\psi_j}) B \bigr)$, and its variance by
$ \s_{j}^{2} = \tr \bigl( T(\ketbra{\psi_j}) B^2 \bigr) - 
{\tr}^2 \bigl( T(\ketbra{\psi_j}) B \bigr)$.
In terms of
$\theta_j$, this reduces to 
$b_j =\inp{\theta_j}{B \theta_j}$ 
and
$\s_{j}^{2} =\inp{\theta_j}{B^2 \theta_j} - \inp{\theta_j}{B \theta_j}^2$. 
Similarly, the l.h.s. of
\ref{margel} equals 
$| \overline{\a_0}\a_1 \inp{\theta_0}{A\theta_1} + \a_0\overline{\a_1}
\inp{\theta_1}{A\theta_0} |$, a quantity bounded by $| \inp{\theta_0}{A\theta_1} |$
since $2|\a_0|\cdot|\a_1| \le 1$. Formula \ref{margel} is then a direct application
of Lemma \ref{lem4}.

To reduce the general case to the case above, we note that
a non-pure state $\tau$ can always be represented as a
vector state. Explicitly, suppose that $\tau$ decomposes as 
$\tau = \sum_{i \in \NN} |\beta_i|^2 \ketbra{\ph_i}$. Then define the Hilbert space 
$\tilde{\K} := \bigoplus_{i \in \NN} \K_i$, where each $\K_i$ is a copy of $\K$.
Now since 
$(\bigoplus_{i \in \NN} \K_i) \otimes \H \cong \bigoplus_{i \in \NN} (\K_i
\ten \H)$,  
we may define, for each $X \in \B(\K\ten\H)$, the operator  
$\tilde{X} \in \B(\tilde{\K}\ten\H)$ by 
diagonal action on the components of the sum, i.e. 
$\tilde{X} \bigr( \bigoplus_{i \in \NN}(k_i \ten h_i)\bigr) := 
\bigoplus_{i \in \NN} X(k_i \ten h_i)$. 
If we now define the vector $\tilde{\ph} \in \tilde{\K}$ by 
$\tilde{\ph} = \bigoplus_i \b_i \ph_i$, then we have
for all $X \in \K\ten\H$ and $\chi \in \H$:
\begin{eqnarray*}
\tr \bigl( \tilde{U} ( \ketbra{\tilde{\ph}} \otimes \ketbra{\chi} ) \tilde{U}^*
\tilde{X}\bigr)
&=&
\inp{{\textstyle \bigoplus}_{i \in \NN}(\b_i\ph_i \ten \chi)}
{\tilde{U}^*\tilde{X}\tilde{U}{\textstyle \bigoplus}_{j \in \NN}(\b_j\ph_j \ten \chi)}_{\tilde{\K}\ten \H}\\
&=&
\inp{{\textstyle \bigoplus}_{i \in \NN} (\b_i\ph_i \ten \chi)}{{\textstyle
\bigoplus}_{j \in \NN} U^*XU(\b_j\ph_j \ten
\chi)}_{\tilde{\K}\ten \H}\\
&=&
\sum_{i \in \NN} |\b_i|^2 \inp{(\ph_i \ten \chi)}{U^*XU(\ph_i \ten
\chi)}_{\K\ten\H}\\
&=&
\sum_{i \in \NN} |\b_i|^2 \tr \big( U (\ketbra{\ph_i} \ten \ketbra{\chi})
U^* X \big)
\\
&=&
\tr \bigl( U (\tau \otimes \ketbra{\chi}) U^*X\bigr)
\end{eqnarray*}
The second step is due to the diagonal action of the operators on
$\tilde{\K}\ten\H$.
The problem is now reduced to the vector-case by
applying the above to $\chi = \psi$, $\chi = \psi_0$ or
$\chi = \psi_1$ and on the other hand $X = A$, $X = B$ or $X = B^2$.
\qed
\noindent The backbone of Theorem \ref{cor5} is 
formed by the special  case $\s_0 = \s_1 =0$, 
$[A,B] = 0$ and $\tau = \ketbra{\phi}$, which allows for a short and
transparent proof. 

In order to arrive at a physical interpretation of Theorem \ref{cor5}, we 
focus on the case $B = \tilde{B} \ten \one$,
when information is transferred from $\H$ to $\K$. Indeed, examining $\K$ at time
$t$ yields information about $\H$ at time 0.

\subsubsection{Quality of Information Transfer}\noindent
A small ratio $\frac{\s_0 + \s_1}{| b_0 - b_1 |}$ indicates a good quality of
information transfer.
The ratio equals 0 in the perfect case, when $\sigma_0 = \sigma_1 =
0$. Thus $\tilde{B} \ten \one$ takes a definite value of either $b_0$
or $b_1$,  depending on whether the initial state of $\H$ was $\ket{\psi_0}$
or $\ket{\psi_1}$. 
In this case, one can infer the initial state of $\H$
with certainty by inspecting only the ancilla $\K$. 
More generally, 
it is still possible to reliably determine from 
the ancilla $\K$ whether the open system $\H$ was initially in state
$\ket{\psi_0}$ or $\ket{\psi_1}$ 
as long as the standard deviations are
small compared to the difference in mean, 
$\sigma_0 , \sigma_1 \ll |b_0 - b_1|$. 
\begin{center}
\setlength{\unitlength}{0.9 cm}
\begin{picture}(8,4.5) 
\put(0,1){\line(1,0){8}}
\put(7.2,0.7){\mbox{\footnotesize $b$}}
\put(4,0.6){\line(0,1){3.9}}
\put(4.1,4){\mbox{\footnotesize $p(b)$}}
\qbezier(0,1.2)(0.7,1.3)(1,2.5)
\qbezier(1,2.5)(1.5,5)(2,2.5)
\qbezier(2,2.5)(2.3,1.3)(3,1.2)
\qbezier(3,1.2)(3.03 , 1.205)(8,1.05)
\qbezier(8 ,1.2)(7.3,1.3)(7,2.5)
\qbezier(7,2.5)(6.5,5)(6,2.5)
\qbezier(6,2.5)(5.7,1.3)(5,1.2)
\qbezier(5,1.2)(4.97 , 1.201)(0,1.05)
\put(1.5,2.5){\vector(1,0){0.5}}
\put(1.4,2.7){\mbox{\footnotesize $\sigma_0$}}
\put(1.5,2.5){\vector(-1,0){0.5}}
\put(6.5,2.5){\vector(1,0){0.5}}
\put(6.4,2.7){\mbox{\footnotesize $\sigma_1$}}
\put(6.5,2.5){\vector(-1,0){0.5}}
\put(4,1.5){\vector(1,0){2.5}}
\put(4.2,1.7){\mbox{\footnotesize $| b_1 - b_0 |$} }
\put(4,1.5){\vector(-1,0){2.5}}
\put(0,0.2){\mbox{\scriptsize Probability densities $p$ of $B$ according to
input $\ket{\psi_0}$ and $\ket{\psi_1}$}}
\end{picture}
\end{center}
As the ratio increases, the restriction \ref{margel}
gets less severe, reaching triviality at 
$\s_0 + \s_1 = 2|b_0 - b_1|$.

\subsubsection{Decoherence on the Commutant of the Pointer}\noindent
Assume perfect information transfer, i.e. $\s_0 = \s_1 =
0$. If $[A,B] = 0$, then Theorem \ref{cor5} 
says that 
coherent and mixed initial states yield identical distributions of $A$
at time $t$. 
In order to distinguish, at time $t$, whether or not $\H$ was in a pure state 
at time 0, we will have to use observables $A$ which do not commute
with $B$. But then $A$ and $B$ cannot be observed simultaneously. Summarizing:
\begin{quote}\label{quote2}
\emph{At time $t$, it is possible to distinguish whether $\H$ was in state
$\psi_0$ or $\psi_1$ at time 0. It is also possible to distinguish  whether $\H$ was in state
$\psi$ or $\Col$ at time 0. But it is {\it not} possible to do both.
}
\end{quote}
We emphasize that this holds even when one has all observables of 
the entire closed system $\K \ten \H$ at one's disposal.

\subsubsection{Decoherence on the Open System}\noindent
We consider the final state of the open system $\H$, obtained from the
final state of the closed system $\K \ten \H$ by tracing out the degrees of
freedom of the ancilla $\K$: an initial state $\rho \in S(\H)$
yields final state $\tr_{\K}(T(\rho)) \in S(\H)$.

Suppose that information is transferred to a pointer $B = \tilde{B} \ten \one$ in the
ancilla $\K$ with perfect quality, $\s_0 = \s_1 = 0$. 
Since $[\one\ten \tilde{A}, \tilde{B}\ten\one] = 0$, we see 
from Theorem \ref{cor5} that we have 
$\tr ( T(\ketbra{\psi}) (\one\ten \tilde{A}) ) $ $=$ 
$\tr ( T ( \Col ) ( \one\ten \tilde{A} ) )$
for {\it all} $\tilde{A} \in \B(\H)$,
or equivalently
\begin{equation}\label{schildpad}
   \tr_{\K} \bigl( T(\ketbra{\psi}) \bigr) =
   \tr_{\K} \Bigl( T \bigl( \Col  \bigr) \Bigr) \; .
\end{equation}
In words:
\begin{quote}\emph{Suppose that at time $t$, by making a hypothetical 
measurement of $\tilde{B}$ on the ancilla,
it would be possible to distinguish perfectly
whether the open system had been in state $\psi_0$ or $\psi_1$ at time 0.
Then, by looking only at the observables of the open system, 
it is not possible to distinguish whether $\H$ 
had been in
the pure state $\psi=\a_0\psi_0+\a_1\psi_1$ or the collapsed state
$|\a_0|^2\ketbra{\psi_0}+|\a_1|^2\ketbra{\psi_1}$ at time 0.}
\end{quote}
This statement holds true, regardless whether $\tilde{B}$
is actually measured or not.
(So we do not assume here that such a measurement is physically possible.)
We have shown that the map
$M_0 + M_1 = \tr_{\K} \circ T$,
with $T : \S(\H) \to \S(\K\ten\H)$ the
information-transfer operation defined by
$ T(\rho) := U(\tau\ten\rho) U^*$,
constitutes a physical realization of  
the
abstract decoherence mapping $(M_0 + M_1)$ of section \ref{perse}.

All in all, 
we have proven that decoherence is an unavoidable consequence of
information transfer out of an open system.

\subsubsection{Example}\label{spinflip}\noindent
The simplest possible example of 
unitary information transfer
is the following.
Let $\K\sim\H\sim\CC^2$ be the Hilbert space of a qubit;
let $\psi_0=(1,0)$ and $\psi_1=(0,1)$ be the `computational basis',
and let $U:\CC^2 \otimes \CC^2\to\CC^2 \otimes \CC^2$ 
be the `controlled-not gate'.
Explicitly, $U$ is defined by 
$U\ket{\psi_1 \ten \psi_1} = \ket{\psi_0 \ten \psi_1}$,
$U\ket{\psi_0 \ten \psi_1} = \ket{\psi_1 \ten \psi_1}$,
$U\ket{\psi_1 \ten \psi_0} = \ket{\psi_1 \ten \psi_0}$, and
$U\ket{\psi_0 \ten \psi_0} = \ket{\psi_0 \ten \psi_0}$.
That is, it flips the first qubit whenever the second qubit is set to 1.
Let $\tau$ be the 0 state of the first qubit.

Since the initial state of the second qubit can be read off from the first, 
this situation satisfies the hypotheses of Theorem \ref{cor5} with 
$B = \s_z \ten \one$ and
$\s_0 = \s_1 = 0$.
We verify equation \ref{schildpad}. For any state 
$\ket{\psi} = \a_0 \ket{\psi_0} + \a_1 \ket{\psi_1}$:
\begin{eqnarray*}
U \ket{\psi_0 \ten \psi} &=& \a_0 \ket{\psi_0 \ten \psi_0} 
+ \a_1 \ket{\psi_1 \ten \psi_1} \, := \, \ket{\theta} ;\\
\tr_{\K} \left(\ketbra{\theta}\right) & = & \Col\,. \\
\end{eqnarray*}
Thus we have
$\tr_{\K} \bigl( T(\ketbra{\psi}) \bigr) = \Col$. 
This agrees with equation \ref{schildpad}, since one can easily check that
$\tr_{\K}\bigl( T(\Col)\bigr) $ equals $\Col$ as well.

\subsection{Unitary Information Transfer and State Collapse}\label{realcol}\noindent
We have derived that,
in the context of  
information transfer to an ancillary system, the initial states 
$\ketbra{\psi}$
and $|\a_0|^2 \ketbra{\psi_0} + |\a_1|^2
\ketbra{\psi_1}$ lead to the same final state.
This is decoherence.

State collapse is a much stronger statement:  
if outcome `0' is observed, 
then the system will further behave as if its initial state
had been $\psi_0$ instead of $\psi$.
Similarly, if outcome `1' is observed, then the system 
will behave as if its initial state had been $\psi_1$.
Now suppose that we ignore the outcome.
Since `0' happens with probability $|\a_{0}|^2$
and `1' with probability $|\a_{1}|^2$, 
the system will behave as if its initial state had been $|\a_0|^2 \ketbra{\psi_0} + |\a_1|^2
\ketbra{\psi_1}$. 
We see that collapse implies decoherence. 

The converse does not hold however: imagine a Stern-Gerlach experiment, 
in which a beam of particles in a $\s_x$-eigenstate is split in two 
according to spin in the $z$-direction. 
State collapse is the statement that 
one beam consists of particles with positive spin, the other 
of particles with negative spin and that both beams  
have equal intensity.
Decoherence is the statement that both outgoing beams \emph{together}
consist for 50\% of positive-spin particles and for 50\% of negative-spin
particles. The former statement is strictly stronger than the latter, 
and deserves separate investigation.   

We will therefore answer the following question: suppose that we
transfer information
to an ancilla $\K$, and then separate $\K$ from $\H$,
dividing $\H$ into subensembles according to outcome.
What states do we use to describe these subensembles?
 

\subsubsection{Joint Probability Distributions}\noindent
A special case of an observable is an \emph{event} $p$, which 
in quantum mechanics is represented by 
a projection $P$.
The relative frequency of occurrence of $p$ 
is given by $ {\mathbb P}(p=1) = \tr(\rho P)$. 

The projection $\one - P$ is interpreted as `not $p$'.
Furthermore, if a projection $Q$ corresponding to an observable $q$ 
commutes with $P$, then $PQ$ is again a projection.
According to quantum mechanics, $p$ and $q$ can then be observed
simultaneously, and the projection $PQ$ 
is interpreted as the event `$p$ and $q$ are both observed'.

A state $\rho$ therefore induces a joint probability distribution on
$p$ and $q$:
\begin{center}
\begin{tabular}{ccccccc}
$ \tr(\rho PQ)$&$=$&${\mathbb P}(p=1,q=1)$ &\,,\,
& $ {\mathbb P}(p=0,q=1)$&$=$&$ \tr(\rho (\one -P)Q) $ \\
$  \tr(\rho P(\one -Q))$&$=$&${\mathbb P}(p=1,q=0)$ &\,,\,
& $ {\mathbb P}(p=0,q=0) $&$=$&$ \tr(\rho (\one -P)(\one -Q)) $
\end{tabular}
\end{center}  
Particularly relevant is the case in which $\rho$ is a state on 
a combined space $\K \ten \H$, and
the projections are of the form $Q \ten \one$ and $\one \ten P$.
(The commuting projections are properties of different systems.)
We then have 
${\mathbb P}(p=1,q=1) = 
\tr( (\one \ten P) (Q \ten \one) \rho ) =
\tr \left( P \, \tr_{\K} ((Q\ten\one) \rho)  \right)$.
This holds for all projections $P$ on $\H$, so that
the normalized version of 
$\tr_{\K} ((Q\ten\one) \rho) \in \S(\H)$ 
must be interpreted as 
the state of $\H$, given that $q = 1$.
Similarly, the normalized version of $\tr_{\K} ((\one - Q)\ten\one \rho) \in \S(\H)$
is the state of $\H$, given that $q=0$ is observed. 

\subsubsection{Collapse }\noindent
Let $T : \rho \mapsto U (\tau \ten \rho) U^*$ from $\S(\H)$ to $\S(\K \ten
\H)$ be an information transfer from $\H$ to a pointer-projection $Q \in
B(\K)$. That is, $\tr((Q\ten\one) T(\ketbra{\psi_0})) = 0$ and 
$\tr((Q\ten\one) T(\ketbra{\psi_1})) = 1$, so that at time $t$, one can 
see from $\K$ whether $\H$ was in state $\psi_0$ or $\psi_1$ at time $0$.

Since $Q \ten \one$ 
commutes with all of $\one \ten B(\H)$, it is possible to separate
$\H$ from $\K$, and divide $\H$ into subensembles according to 
the outcome of $Q$.
This is done as follows: 
with any measurement 
on $\H$, a simultaneous measurement of
$Q$ on $\K$ is performed to determine in which ensemble 
this particular system should fall. 
It follows from the above that the $1$-ensemble should be
described by the normalized version of  
$M_1 (\rho) := \tr_{\K} ((Q\ten \one) T(\rho))$, and the $0$-ensemble by
the normalized version of $M_0(\rho) := \tr_{\K} ((\one-Q\ten \one) T(\rho))$.
Since $Q$ commutes with $B(\H)$,
this is just conditioning on a classical probability space
at time $t$.
We have arrived at an interpretation of the map 
$M(\rho) := M_0(\rho) \oplus M_1(\rho)$ of section \ref{perse}.

We will now prove that $M$ takes the form
$M(\ketbra{\psi}) = |\a_0|^2 \tr_{\K}T(\ketbra{\psi_0}) \oplus 
|\a_1|^2 \tr_{\K}T(\ketbra{\psi_1})$.
This is a strong physical statement. For instance,
any spin-system 
$\a_0 \ket{\psi_0} + \a_1 \ket{\psi_1}$ that is
found to have spin 1 in the $z$-direction 
may subsequently be treated as if it had been in state $\psi_1$ 
at time 0. This is nontrivial: a priori, it is perfectly conceivable
that the different initial states $\psi_0$ and $\psi$ result
in different final states, even though they yield the same $Q$-output.  

One could alternatively, (and more traditionally), arrive at the `collapse of the wavefunction'
$M(\ketbra{\psi}) = |\a_0|^2 \tr_{\K}T(\ketbra{\psi_0}) \oplus 
|\a_1|^2 \tr_{\K}T(\ketbra{\psi_1})$ by assuming that, 
at time 0, 
the quantum system 
makes either the jump $\ketbra{\psi} \mapsto \ketbra{\psi_0}$ 
or the jump $\ketbra{\psi} \mapsto \ketbra{\psi_1}$.
Since we arrive at the same conclusion, namely the above 
`collapse of the wavefunction', using only 
open systems,
unitary transformations and
the probabilistic
interpretation of quantum mechanics, such an  
assumption of `jumps' at time $0$ is made redundant.

\begin{proposition}\label{prop6}
Let $T : \rho \mapsto U (\tau\ten\rho) U^*$ from $\S(\H)$ to $\S(\K \ten
\H)$ satisfy $\tr((Q\ten\one) T(\ketbra{\psi_0})) = 0$ and
$\tr((Q\ten\one) T(\ketbra{\psi_1})) = 1$ for some `pointer-projection' $Q$
on $\K$. 
Define a map $M : \S(\H) \mapsto \S(\H)\oplus \S(\H)$ by 
$M(\rho) := \tr_{\K} ((\one -Q\ten\one) T(\rho)) \oplus \tr_{\K} ((Q\ten\one)
T(\rho))$.
Then for $\psi = \a_0\psi_0 + \a_1 \psi_1$ we have
$M(\ketbra{\psi}) = |\a_0|^2 \tr_{\K}T(\ketbra{\psi_0})
\oplus |\a_1|^2 \tr_{\K}T(\ketbra{\psi_1})$.
\end{proposition}
This can be seen almost directly from Proposition \ref{abscol}:

\proof 
Since $M_{1}(\ketbra{\psi_0}) \geq 0$ is a positive operator, we may
conclude from $\tr \left(M_{1}(\ketbra{\psi_0})\right) = 0$ that
$M_{1}(\ketbra{\psi_0}) = 0$. Similarly $M_{0}(\ketbra{\psi_1}) = 0$.
Utilizing Proposition \ref{abscol}, we find that
$M(\ketbra{\psi}) = |\a_0|^2 M_{0}(\ketbra{\psi_0}) 
\oplus |\a_1|^2 M_{1}(\ketbra{\psi_1})$.
The proof is completed by noting from $\tr_{\K}((\one - Q\ten \one) T(\ketbra{\psi_1}))=0$ 
that 
$\tr_{\K}(T(\ketbra{\psi_1})) =   
\tr_{\K}((Q\ten \one) T(\ketbra{\psi_1})) =
M_{1}(\ketbra{\psi_1})$,
and similarly that
$\tr_{\K}(T(\ketbra{\psi_0})) =   M_{0}(\ketbra{\psi_0})$.
\qed
\noindent We summarize: 
\begin{quote}
\emph{Consider an ensemble of systems of type $\H$ in state $\psi$. 
	Suppose that information is transferred   
	to a pointer-projection 
	$Q$ on an ancillary system $\K$. 
	Subsequently, the ensemble is divided into two subensembles according
	to outcome. 
	Then all observations on $\H$ made afterwards, conditioned on the 
	observation that
	the measurement outcome was 0, will be as if the system had 
	originally been in 
	the collapsed state $\psi_0$ instead of $\psi$.
	No independent `collapse postulate' is needed to arrive at this
	conclusion.}
	
\end{quote}

\subsubsection{Example}\noindent 
In the simple model of information transfer introduced in 
Section~\ref{ITSC}, we will now demonstrate why 
repeated spin-measurements 
yield identical outcomes. 

The probed system is once again a single spin $\H = \CC^2$,
whereas the ancillary system now consists of two spins, 
$\K = \CC^2 \otimes \CC^2$ in initial state $\ket{\psi_0 \ten \psi_0}$. 
Repeated information-transfer, first to pointer 
$\s_{z,1}$ and then to $\s_{z,2}$, is then represented by the unitary 
$U := U_2 U_1$ on $\K \ten \H$. In this expression, $U_1$ is the 
controlled not-gate flipping the first qubit of $\K$ if $\H$ is 
set to 1,
and $U_2$ flips the second qubit of $\K$ if $\H$ is set to 1. 

Since $U  \ket{\psi_0 \ten \psi_0 \ten (\a_0\psi_0 + \a_1\psi_1)} = 
\ket{\a_0 \psi_0 \ten \psi_0 \ten \psi_0}
+ \ket{\a_1 \psi_1 \ten \psi_1 \ten \psi_1}$,
we can explicitly calculate the joint probability distribution 
on the two pointers $\s_{z,1}$ and $\s_{z,2}$ in the final state: 
\begin{center}
\begin{tabular}{rcccccl}
$ {\mathbb P}(s_{z,1}= \,\,\,\,\,1,s_{z,2}=1)$ & $=$ & $|\a_1|^2$ &\,,\,&  $0$ &
$=$ & ${\mathbb P}(s_{z,1}=\,\,\,\,\,1,s_{z,2}=-1)$ \\
$ {\mathbb P}(s_{z,1}=-1,s_{z,2}=1)$ & $=$ & $0$ &\,,\,& $|\a_0|^2$&$=$&
${\mathbb P}(s_{z,1}=-1,s_{z,2}=-1)$
\end{tabular}
\end{center}
In particular, we see that if the first outcome is $1$ (which happens with
probability $|\a_1|^2$), then so is the second.
Proposition \ref{prop6} shows that this is the general situation, 
independent of the (rather simplistic) details of this particular model. 

\subsection {Information Leakage to the Environment}\label{leakage}\noindent  
On closed systems decoherence does not occur, because unitary time
evolution preserves the purity of states.  
However, macroscopic systems are almost never closed.

Imagine, for example, that $\H = { \mathbb C}^2$ represents a two-level atom, 
and $\K$ some large measuring device.
Information about the energy $ \one \ten \sigma_z$ of the atom is
transferred to the apparatus, where it is stored as the position 
$\tilde{B} \ten \one$ of a pointer. 
Then as soon as information on the pointer-position $ \tilde{B} \ten \one$
leaves the system, collapse on the combined atom-apparatus system takes place. 
For example, a ray of light may reflect on the pointer, revealing its
position to the outside world. (See \cite{JandZ}.)
It is of course immaterial whether or not someone is actually {\it looking} at the photons. 
If even the smallest speck of light were to fall on the pointer, the information
about the pointer position would 
already be encoded in the light, causing full collapse on the atom-apparatus system.
(See \cite{Zur} for an example.)

The quality of this information transfer will not be perfect.
If a macroscopic system is interacting normally with the outside world, (the occasional photon happens to 
scatter on it, for instance),
then a number of macroscopic observables $X$ will leak
information continually, with a macroscopic uncertainty $\sigma$. 
This enables us to apply Theorem \ref{cor5}. 
It says that all coherences between eigenstates $\psi_{x_1}$ and $\psi_{x_2}$ of 
macroscopic observables 
$X$ are continually vanishing on the macroscopic system $\L$,  
provided that their eigenvalues $x_1$ and $x_2$ satisfy $| x_1 - x_2 | \gg 2\sigma$. 
(The pointer, e.g. a
beam of light, is outside the system, so that $\delta = 0$.)

Take for example a collection of $N$ spins, $\L = \bigotimes_{i = 1}^{N}
{\mathbb C}^2$.
Suppose that for $\alpha = x,y,z$, the average spin-observables $S_{\alpha} = \frac{1}{N}\sum_{i = 1}^{N} \sigma_{\alpha}^{i}$ 
are continually being measured
with an accuracy\footnote{Since $[S_{x}, S_{y}] \neq 0$, they cannot be
simultaneously measured with complete accuracy, see e.g.\cite{Wer}.
However, this problem disappears if the accuracy satisfies 
$\s^2 \geq \frac{1}{2} \|[S_{x} , S_{y}]\| = \frac{1}{N}$, see \cite{scriptie}.
For large $N$, (typically $N \sim 6 \times 10^{23}$),
this allows for extremely accurate measurement. 
} $N^{-\frac{1}{2}} \ll \sigma \ll 1$. Then between 
macroscopically different eigenstates of $S_{\alpha}$, i.e. states for which the eigenvalues
satisfy 
$|s_{\alpha} - s'_{\alpha}| \gg \sigma$, coherences 
are constantly disappearing. 
However, the information leakage need not have any effect 
on states which only differ on a microscopic scale.
Take for instance $\rho \otimes \ketbra{+} $ 
and $\rho \otimes \ketbra{-} $, with $\rho$ an arbitrary state on $N-1$ spins.
Indeed, $|s_{\alpha} - s'_{\alpha}| \leq 2/N \ll \sigma$, so 
Theorem \ref{cor5} is 
vacuous in this case: no decoherence occurs. 

We see how the variance $\sigma^2$ produces a smooth boundary between the macroscopic 
and the microscopic world: macroscopically distinguishable states
(involving $S_{\alpha}$-differences $\gg \sigma$) continually 
suffer from loss of coherence, while states that only differ microscopically 
(involving $S_{\alpha}$-differences $\ll \sigma$) 
are unaffected. 

In case of a system monitored by a macroscopic measurement apparatus, we are 
interested in coherence between eigenstates of the macroscopic
pointer. By definition, these eigenstates are macroscopically distinguishable. 
We may then give the following answer 
to the question why
it is so hard, in practice, to witness coherence:  
\begin{quote}
\emph{If information leaks from the pointer into the outside world,
decoherence 
takes place on the 
combination of system and measurement apparatus. In practice, macroscopic
pointers constantly leak information.}
\end{quote}

\section{Closed Systems}\label{oester}\noindent
Closed systems evolve according to unitary time evolution, so that
coherence which is present initially will still be there at later times. 
Yet on macroscopic systems, coherent superpositions are almost never observed. 
Why is this the case? 

\subsection {Macroscopic Systems} \label{macro}\noindent
Because of the direct link that it provides between
the scale of a system on the one hand, and on the other hand
the difficulties in witnessing coherence, we feel that 
the following line of reasoning, essentially due to Hepp \cite{Hepp},
is the most important mechanism hiding coherence.

Let us first define what we mean by macroscopic and microscopic
observables. We consider a system consisting of $N$ distinct
subsystems, i.e. $\K = \bigotimes_{i=1}^{N} \K_i$.
If one thinks of $\K_i$ as the atoms out of which a macroscopic system
$\K$ is constructed, $N$ may well be in the order of $10^{23}$.

We will define the \emph{microscopic} observables to be the ones that refer
only to one particular subsystem $\K_i$:  
\definition
An observable $X \in B(\K)$ is called \emph{microscopic} if it is of the
form $X = \one \otimes \ldots \otimes \one \otimes X_i \otimes \one 
\otimes \ldots \otimes \one$ for some $i \in \{ 1,2,\ldots,N \}$
and some $X_i \in B(\K_i)$. \bigbreak
\noindent In this situation we will identify $X_i \in B(\K_i)$ with $X \in B(\K)$. 
We take macroscopic
observables to be averages of microscopic observables `of the same size':
\definition
An observable $Y \in B(\K)$ is called \emph{macroscopic} if it is of the
form $Y = \frac{1}{N} \sum_{i=1}^{N} Y_i$, 
with $Y_i \in B(\K_i)$ such that $\|Y_i\| \leq \|Y\|$. \bigbreak
\noindent We will only use the term `macroscopic' in this narrow
sense from here on, even though there do exist observables which are 
called `macroscopic' in daily life, but do not fall under the above
definition.

Now suppose that we transfer information from a system $\H$ to a 
macroscopic system $\K = \bigotimes_{i=1}^{N} \K_i$, using a 
macroscopic pointer $\tilde{B} \in B(\K)$. As explained before, we then have a map
$T : \S(\H) \rightarrow \S(\K \otimes \H)$ such that the pointer
$\tilde{B} \otimes \one$ has 
different expectation values $b_0$ and $b_1$ in
the states
$T(\ketbra{\psi_0})$ and $T(\ketbra{\psi_1})$.

Since $\tilde{B}$ is macroscopic, it is unrealistic to require 
$T(\ketbra{\psi_0})$ and $T(\ketbra{\psi_1})$
to be eigenstates of $\tilde{B}$. Instead, we will require their standard 
deviations in $\tilde{B}$ to be negligible compared to their difference in mean,
i.e. $\s_0 \ll |b_0 - b_1|$ and $\s_1 \ll |b_0 - b_1|$.  

After this information transfer, we try to distinguish whether the 
system $\H$ had initially been in the coherent state 
$\a_0 \ket{\psi_0} + \a_1 \ket{\psi_1}$
or in the incoherent mixture $|\a_0|^2 \ketbra{\psi_0} + |\a_1|^2 \ketbra{\psi_1}$. 
We have already shown that this cannot be done by measuring 
observables in $\one \otimes B(\H)$. 
The following adaptation of Theorem \ref{cor5} shows that it 
is also impossible to do this
by measuring
macroscopic or microscopic observables on the closed system $\K \otimes \H$.

\begin{corollary}\label{cor6}
Let $\psi_0$ and $\psi_1$ be orthogonal unit vectors in a Hilbert space $\H$
and let $\tau \in \S(\K)$ be a state on the Hilbert space 
$\K = \bigotimes_{i=1}^{N} \K_i$.
Let $U : \K \ten \H \rightarrow \K \ten \H$ be unitary and define
$T : \S(\H) \rightarrow \S(\K \ten \H)$ by
$T(\rho) = U (\tau \ten \rho) U^* $.
Let $\tilde{B}$ be a macroscopic observable in 
$B(\K)$, and define $B := \tilde{B} \ten \one$. 
Denote by $b_j$ 
and $\s_{j}^{2}$  its expected value  and variance 
in the state $T(\ketbra{\psi_j})$ for $j = 0,1$.
Suppose that $b_0 \neq b_1$. 
Then for all $\psi=\a_0\psi_0+\a_1\psi_1$ with $|\a_{0}|^2 + |\a_{1}|^2 = 1$
and for all microscopic and macroscopic
observables $A \in B(\K \otimes \H)$, we have
$$ \label{mac}
   \Bigl| \tr \bigl( T(\ketbra{\psi}) A \bigr) -
   \tr \Bigl( T\bigl( |\a_0|^2 \ketbra{\psi_0} + |\a_1^2| \ketbra{\psi_1}\bigr) A \Bigr) \Bigr| 
   \le
   {{\frac{2}{N} \norm B+\s_0+\s_1}\over{|b_0-b_1|}}\norm A \,.
$$
\end{corollary}
\proof 
If A is microscopic, we have $\|[A,B]\| = \|[A_i , \frac{1}{N}
\sum_{j=1}^{N} B_j]\| = \frac{1}{N} \|[A_i, B_i]\| \leq \frac{2\|A\|
\|B\|}{N}$.
If A is macroscopic, we have $\|[A,B]\| = \|[\frac{1}{N+1}
\sum_{i=0}^{N} A_i , \frac{1}{N} \sum_{j=1}^{N} B_j]\| = 
\frac{1}{N(N+1)} \sum_{i=1}^{N} \|[A_i, B_i]\| \leq \frac{2\|A\|
\|B\|}{N}$.
Either way, we can now apply Theorem \ref{cor5}. 
\qed

\subsection{Examples}\noindent
In order to illustrate the above,  
we discuss four examples of information transfer to
a macroscopic system.

\subsubsection{The Finite Spin-Chain}\label{spin-chain}\noindent
We study a single spin $\H = \CC^2$ in interaction with 
a large but finite spin-chain $\K = \bigotimes_{i=1}^{N} \CC^2$,
the latter acting as a measurement apparatus.
Once again, let $\psi_0=(1,0)$ and $\psi_1=(0,1)$ be the `computational
basis'. Initially, all spins in the spin-chain are down: 
$\tau = \ketbra{\psi_0 \otimes\ldots \otimes \psi_0}$.
Let $U_i:\K\ten\H\to\K\ten\H$ be the `controlled-not gate',
which flips spin number $i$ in the chain whenever the single qubit is set to 1.
(We define $U_j = \one$ for $j \notin \{1,2,\ldots,N\}$.)
$$U_i = \one \otimes P_{-} + \s_{x,i} \otimes P_{+} \quad \mbox{with} \quad
	P_+ =  \left( \matrix{	1&0 \cr
				0&0 \cr}
		\right), \quad
	P_- =  \left( \matrix{	0&0 \cr
				0&1 \cr}
		\right)	\,.
$$
In discrete time $n \in {\Bbb Z}$, the unitary evolution is given 
by $n \mapsto U_n U_{n-1} \ldots U_2 U_1$. (See \cite{Hepp}.)
This represents a single spin flying over a spin-chain from 1 to $N$,
interacting with spin $n$ at time $n$. 

Obviously
$U_N  \ket{\psi_0 \otimes\ldots\otimes\psi_0} \otimes \ket{\psi_0}=
     \ket{\psi_0 \otimes\ldots\otimes\psi_0} \otimes \ket{\psi_0} $
and     
$U_N \ket{\psi_0 \otimes\ldots\otimes\psi_0} \otimes \ket{\psi_1}=
    \ket{\psi_1 \otimes\ldots\otimes\psi_1} \otimes \ket{\psi_1}$.
We consider the average spin of the spin-chain as pointer, 
$B = \frac{1}{N} \sum_{i=1}^{N} \s_{z,i}$.
This makes the map $T : \rho \mapsto U_N \tau \otimes \rho U_N^{*}$ an
information transfer to a macroscopic system.
Applying Corollary \ref{cor6} with $b_0 = -1$, $b_1 = 1$ and $\s_0 = \s_1 = 0$
yields the estimate
$$
   \Bigl| \tr \bigl( T(\ketbra{\a_0\psi_0+\a_1\psi_1}) A \bigr) -
   \tr \Bigl( T\bigl( |\a_0|^2 \ketbra{\psi_0} + |\a_1^2| \ketbra{\psi_1}\bigr) A \Bigr) \Bigr| 
   \le
   \frac{1}{N} \norm A
$$
for all microscopic and macroscopic $A \in B(\K \ten \H)$. 
Indeed, in this particular model, the estimated quantity is identically
zero since $\inp{\psi_0 \otimes \ldots \psi_0}
{X_i \,\psi_1 \otimes\ldots\otimes\psi_1 } = 
\inp{\psi_0}{\psi_1}^{N-1}\inp{\psi_0}{X_i \psi_1} = 0$ for all 
microscopic $X_i$.

Of course coherence \emph{can} be detected on the closed system $\K \ten \H$,
but only using observables that are neither macroscopic nor microscopic, 
such as $\s_x \ten \ldots \ten \s_x$.
\subsubsection{Finite Spin-Chain at Nonzero Temperature}\noindent
A more realistic initial state for the spin-chain is the
nonzero-temperature state $\tau_{\b} = \frac{e^{-\beta H}}{\tr e^{-\beta H}}$.
For the spin-chain Hamiltonian we will take 
$H = \sum_{i} \s_{z,i} = NB$, so that $\tau_{\b}$ 
becomes the tensor product of
$N$ copies of the $\CC^2$-state 
$$\hat{\tau}_{\b} = \frac{1}{e^{\beta} + e^{-\beta}}  
		\left( \matrix{	e^{-\beta}	&0 \cr
				0		&e^{\beta} \cr}
		\right)\,.
$$
With the same time-evolution as before, we have
$T(\ketbra{\psi_0}) = \ketbra{\psi_0} \otimes \r_{\b}$ and 
$T(\ketbra{\psi_1}) = \ketbra{\psi_1} \otimes \r_{-\b}$. 
Again we choose the mean energy $B$ as our pointer.
\mbox{A brief} calculation shows that 
$\tr(B\tau_\b) = \frac{e^{-\beta} - e^{\beta}}{e^{\beta} +
e^{-\beta}} =: \e(\beta)$ and that
$\tr(B^2 \tau_\b) - \tr(B \rho_\b)^2 = \frac{1}{N}(1 - \e^2(\b)) $.
Corollary \ref{cor6} now gives us, for microscopic and macroscopic $A$,
$$
   \Bigl| \tr \bigl( T(\ketbra{\a_0\psi_0+\a_1\psi_1}) A \bigr) -
   \tr \Bigl( T\bigl( |\a_0|^2 \ketbra{\psi_0} + |\a_1^2| \ketbra{\psi_1}\bigr) A \Bigr) \Bigr| 
   \le
   \left( \frac{1}{\e(\b) N} + \frac{\sqrt{1 - \e^2(\b)}}{\e(\b) \sqrt{N}}\right) 
   \norm A \,.
$$
For large $N$, we see that the term $\sim \frac{1}{N}$ due to
the fact that $[A,B] \neq 0$ is dominated by the thermodynamical
fluctuations, which of course go as $\sim \frac{1}{\sqrt{N}}$. 
In statistical physics, it is standard practice to neglect even the latter.

\subsubsection{Energy as a Pointer}\noindent
Hamiltonians often fail to be macroscopic in our narrow sense of the word,
since they are generically unbounded and contain interaction terms.
However, this does not imply failure of our scheme to estimate coherence.

    
For example, consider an $N$-particle system with Hilbert space  $\K =
\bigotimes_{i=1}^{N} \K_{i}$ 
and Hamiltonian $H = \sum_{i=1}^{N} \frac{p_{i}^{2}}{2m_{i}} + V(x_1, x_2,
\ldots, x_{N} )$.   
Information is transferred from $\H$ to $\K$ with $H$ as pointer, that is 
the two states $\tr_{\H}(T(\ketbra{\psi_0}))$ and
$\tr_{\H}(T(\ketbra{\psi_1}))$ have
different energies $E$ and $E'$.
Without loss of generality, assume that they are vectorstates:
$\tr_{\H}(T(\ketbra{\psi_0})) = \ketbra{\psi}$ and
$\tr_{\H}(T(\ketbra{\psi_1})) =
\ketbra{\psi'}$. (Density matrices can always be represented as vectors
on a different Hilbert space, cf. the proof of Theorem \ref{cor5}.)

We thus have two vector states $\ket{\psi}$ and $\ket{\psi'}$ with different
energies $E := \inp{\psi}{H\psi}$ and $E' := \inp{\psi'}{H\psi'}$. 
We estimate the coherence between $\ket{\psi}$ and $\ket{\psi'}$ on $x_{n}$, 
the position of particle $n$.
\begin{eqnarray*}
(E - E')\inp{\psi}{x_{n} \psi'} &=& \inp{E \psi}{x_{n} \psi'} - \inp{x_{n}\psi}
{E' \psi'}\\
&=& \inp{H \psi - (H - E) \psi}{x_{n} \psi'} - \inp{x_{n} \psi}{ H \psi' - (H
- E')\psi'}\\
&=& \inp{[H,x_{n}] \psi}{ \psi'} - \inp{(H - E) \psi}{x_{n} \psi'} + 
\inp{x_{n} \psi}{(H - E')\psi'}\\
\end{eqnarray*}
Now since 
$
[H,x_{n}] = \frac{1}{2m_{n}} [p_{n}^{2}, x_{n}] = \frac{-i\hbar
p_{n}}{m_{n}} 
$, we can apply the Cauchy-Schwarz inequality in each term to obtain
$$
|E - E'| \, |\inp{\psi}{x_{n} \psi'}| \leq 
\frac{\hbar}{m_{n}} \sqrt{\inp{\psi}{p_{n}^{2} \psi}} +
\sqrt{\inp{\psi}{x_{n}^2 \psi}} \sqrt{\inp{\psi'}{(H - E')^2 \psi'}} +
\sqrt{\inp{\psi'}{x_{n}^2 \psi'}} \sqrt{\inp{\psi}{(H - E)^2 \psi}} \;.
$$ 
If we define the characteristic speed
$V_{n} := \sqrt{ \inp{\psi}{(\frac{p_{n}}{m_{n}})^2 \psi} }$,
the characteristic positions $ X_{n} := \sqrt{\inp{\psi}{x_{n}^2 \psi}}$  
and $ X'_{n} := \sqrt{\inp{\psi'}{x_{n}^2 \psi'}}$, and the 
standard deviations $\sigma := \sqrt{\inp{\psi}{(H - E)^2 \psi}}$ and
$\sigma' := \sqrt{\inp{\psi'}{(H - E')^2 \psi'}}$, we obtain
$$
|\inp{\psi}{x_{n} \psi'}| \leq \frac{\hbar V_{n} + \sigma X'_{n} +
\sigma' X_{n}}{|E - E'|}\;.
$$ 
As such, this doesn't tell us very much. We will have to make some
physically plausible assumptions on the state of the system
in order to obtain results.
First, we assume that the
system is encased in an $L \times L \times L$ box so that $X_{n},
X'_{n} \leq L$. Also, we assume $V_{n} < c$. This yields   
$
|\inp{\psi}{x_{n} \psi'}| \leq \frac{\hbar c + L( \sigma + \sigma' )}{|E - E'|}
$. 
Secondly, we assume that scaling the system in any meaningful way will 
produce $|E - E'| \sim N$
and $\sigma + \sigma' \sim  \sqrt{N}$, so that the coherence on $x_n$
approaches zero as $\sim\frac{1}{\sqrt{N}}$.
Notice the almost thermodynamic lack of detail required for this estimate.

\subsubsection{Schr\"odinger's Cat}\noindent
Let us finally analyze the rather drastic extraction of information from a
radioactive particle that has become known\footnote{
Actually, Schr\"odinger's proposal was slightly different.
In the original thought experiment, death of the cat was correlated 
with decay of the atom at time $t$ instead of $0$, which wouldn't
make it an information transfer in our sense of the word. 
} 
as `Schr\"odinger's cat'. (See \cite{Sch}.) 
The experiment is performed as follows. We are interested in a radioactive
particle. Is it in a decayed state $\psi_0$ or in a non-decayed state
$\psi_1$? 

In order to determine this, we set up the following experiment. 
A Geiger counter is placed next to the radioactive particle. If the particle
decays, then the Geiger counter clicks. A mechanism then releases a hammer,
which smashes a vial of hydrocyanic acid, killing a cat.
All of this happens in a closed box no higher than $1 m$, and
completely impenetrable to
information.  
A measurement of the atom is done as follows: 
first, place it inside the box. Then wait for
a period of time that is long compared to the decay time of the atom.
Finally, open the box, 
and inspect whether the cat has dropped dead or is still standing upright. 

The atom is described by a Hilbert space $\H$, the combination of
Geiger counter, mechanism, hammer, vial and cat by a Hilbert space $\K$.
Initially, the latter is prepared in a state $\ket{\theta}$.
As a pointer, we take the center of mass of the cat, 
$Z := \frac{1}{N} \sum_{i=1}^{N} z_i$. In this expression,
$N$ is the amount of atoms out of which the cat is constructed,
and $z_i$ is the $z$-component of particle number $i$.
(It is a harmless assumption that all atoms in the cat have the same mass.)  
Since the 
box only measures $1 m$ in height, we may take $\|Z\| = 1$.
The unitary evolution $U \in B(\K \otimes \H)$ then  produces
$U \ket{\psi_0 \otimes \theta} := \ket{\gamma_0}$ and
$U \ket{\psi_1 \otimes \theta} := \ket{\gamma_1}$, which are 
eigenstates\footnote{
As discussed before, it would be more realistic to allow for a nonzero 
variance $0 < \s_j \ll 1$ instead of requiring $\theta_j$ to be
eigenstates of $Z$. We use $\s_j = 0$ for clarity,
leaving the argument essentially unchanged .   
} of $Z$ with different eigenvalues.

Suppose that, initially, the atom is either in the decayed
state $\psi_0$ with probability $|\a_0|^2$ or in the
non-decayed state $\psi_1$ with probability $|\a_1|^2$.
That is, the initial state is the incoherent mixture 
$|\a_0|^2 \ketbra{\psi_0} + |\a_1|^2 \ketbra{\psi_1}$.
By linearity, the final state is then the incoherent
state $|\a_0|^2 \ketbra{\gamma_0} + |\a_1|^2 \ketbra{\gamma_1}$.

On the other hand,
if the atom starts out in the coherent superposition 
$\a_0 \ket{\psi_0} + \a_1 \ket{\psi_1}$, then 
the combined system ends up in the coherent state
$U \ket{(\a_0 \psi_0 + \a_1 \psi_1) 
\otimes \theta} = \a_0 \ket{\gamma_0} + \a_1 \ket{\gamma_1}$.

The question is now this: why do we not notice the difference 
between these two situations if we open the box?
First of all, according to Theorem \ref{cor5} (and the observations following 
it in section \ref{quote2}), 
it is impossible to detect coherence between $\gamma_0$
and $\gamma_1$ \emph{and} ascertain the position of the cat.
Upon opening the black box, we must make a choice. 

Secondly, according to the discussion in section \ref{leakage}, the 
coherences between the macroscopically different states $\gamma_0$
and $\gamma_1$ are extremely volatile. Any speck of light falling 
on the cat will reveal its position with reasonable accuracy, 
causing the coherence 
to disappear according to Theorem \ref{cor5}.

Yet even if we were able to open the box without any information on the
position of the cat leaking out, even then would we be unable to detect
coherence between $\gamma_0$ and $\gamma_1$.
Apply Corollary \ref{cor6} to the transfer of information from atom to cat.
We have $\s_0 = \s_1 = 0$, and with pointer $Z$ we have $\|Z\| = 1$
(the height of the box is 1 m) and $z_1 - z_0 = 0.1$ (the difference 
between a cat that is standing up and one that has dropped dead is 10
$cm$). We then obtain for all macroscopic and microscopic $A$:
$$
	\left| \inp{\a_0 \gamma_0 + \a_1 \gamma_1}
	{A \a_0 \gamma_0 + \a_1 \gamma_1} -
	\left( |\a_0|^2 \inp{\gamma_0}{A \gamma_0} +
	|\a_1|^2 \inp{\gamma_1}{A \gamma_1} \right) \right|
	\leq \frac{20}{N} \|A\|\;.
$$
On the subset of observables we are normally able to measure,
the distinction between coherent and incoherent mixtures 
practically vanishes for $N \sim 10^{23}$. 
For all practical intents and purposes, it is completely 
harmless to assume that the 
final state of the cat is 
$|\a_0|^2 \ketbra{\gamma_0} + |\a_1|^2 \ketbra{\gamma_1}$
instead of
$\a_0 \ket{\gamma_0} + \a_1 \ket{\gamma_1}$.
But it would be false to state that the former has actually been observed.

\section{Conclusion}\noindent
In open systems, we have proven that decoherence is a necessary consequence
of information transfer to the outside. More in detail, 
we have reached the following conclusions:
\begin{itemize}
\item[-]Suppose that an open system $\H$ interacts with an ancillary system
$\K$ in such a way, that it is possible, in principle, to determine from
$\K$ whether $\H$ had been in state $\psi_0$ or $\psi_1$ before the
interaction. If $\H$ started out in a coherent state 
$\a_0 \ket{\psi_0} + \a_1 \ket{\psi_1}$, then it will behave after
the information transfer as if it had started out in the incoherent mixture
$|\a_0|^2 \ketbra{\psi_0} + |\a_1|^2 \ketbra{\psi_1}$ instead.
This is called `decoherence'.
\item[-]Suppose again that the information whether $\H$ was in state
$\psi_0$ or $\psi_1$ is transported to an ancillary system $\K$.
This is done with an ensemble of $\H$-systems described by the state 
$\a_0 \ket{\psi_0} + \a_1 \ket{\psi_1}$.
The ensemble is then split into subensembles, according to outcome.
The `0-ensemble' then behaves as if it had been in state $\psi_0$
at the beginning of the procedure,
and the `1-ensemble' as if it had started in state $\psi_1$.
This is called `state collapse'.
\item[-]These results were obtained entirely within the framework of 
traditional quantum mechanics and unitary time evolution on a
larger, closed system containing $\H$. No `reduction-postulate' 
is needed. From Proposition \ref{abscol}, we see that any information
extraction causes collapse, quite independent of its particular mechanism.  
\item[-]On the closed system containing the smaller, open one no 
decoherence occurs in principle. 
In practice however, closed systems are very hard to achieve. 
We have argued that 
information transfer from a macroscopic observable $A$, performed with 
macroscopic precision $\s$, causes decoherence between eigenstates 
of $A$ if their values satisfy $\s \ll |a_1 - a_0|$. 
Since information on macroscopic observables tends to leak out, 
coherence between macroscopically different states tends to vanish.   
\end{itemize}
Still, even if the combined system $\K \ten \H$ is considered perfectly
closed, there are some results to be obtained.
Again, we investigated the case that a system $\H$ interacts 
unitarily with a system $\K$ 
in such a way that the information whether $\H$ was in state $\psi_0$
or $\psi_1$ can be read off from a pointer in $\K$.
We have reached the following conclusions concerning the closed system $\K
\ten \H$:
\begin{itemize}
\item[-]Using only observables on the closed system that commute with the pointer, 
it is impossible to detect whether $\H$ had started out in state 
$\a_0 \ket{\psi_0} + \a_1 \ket{\psi_1}$ or
$|\a_0|^2 \ketbra{\psi_0} + |\a_1|^2 \ketbra{\psi_1}$.
Physically, this means that it is impossible to distinguish between 
coherent and incoherent initial states while at the same time 
distinguishing between $\psi_0$ and $\psi_1$.  
\item[-]Suppose that the closed system $\K \ten \H$ is macroscopic,
and that one has access to its macroscopic and microscopic
observables only. Then it is almost impossible to distinguish 
whether $\H$ had started out in state 
$\a_0 \ket{\psi_0} + \a_1 \ket{\psi_1}$ or
$|\a_0|^2 \ketbra{\psi_0} + |\a_1|^2 \ketbra{\psi_1}$.
We have obtained upper bounds on the coherences
$\inp{\psi_0}{A \psi_1}$, evaluated on microscopic or macroscopic $A$.
Assuming perfect information transfer ($\s_0 \!=\! \s_1 \!= 0$),
they approach zero as $\sim \frac{1}{N}$, where $N$ is the size 
of the system.
\end{itemize}
In short: no decoherence ever occurs on perfectly closed systems,
even if they are macroscopic.
It is just very hard to distinguish coherent from incoherent states, 
creating the false impression that it does.

The link between decoherence and macroscopic systems was brought 
forward by Klaus Hepp in his 
fundamental paper \cite{Hepp},
where he considered infinite closed systems,
displaying decoherence in infinite time.
In infinite systems, the microscopic observables form a 
non-commutative C$^*$-algebra $\cal{A}$.
Its weak closure ${\cal A}''$ is considered as the (von Neumann-)algebra of
all observables.
The macroscopic observables form a commutative algebra ${\cal C}$ which is
contained in the centre of ${\cal A}''$, i.e. ${\cal C} \subset {\cal Z} = 
\{ Z \in {\cal A}'' | [Z,A] = 0 \, \, \forall A
\in {\cal A}'' \}$, yet is almost disjoint from the microscopic
observables:
${\cal C} \cap {\cal A} = \CC \one$.  
Transfer of information to a macroscopic observable 
therefore implies perfect decoherence on all microscopic and macroscopic
observables (cf. section \ref{quote2}).

Unfortunately, this transfer cannot be done by any automorphic 
time-evolution, since the
macroscopic observables are central. Hepp proposed information transfer by
a $t \rightarrow \infty$ limit of automorphisms. 
He was able to show that this 
causes decoherence in the weak-operator sense.   
That is, on each fixed microscopic observable, the coherence
becomes arbitrarily small for sufficiently large $t$.   

The paper was criticized by John Bell a few years later \cite{Bell},
on the grounds that, for each fixed time $t$, there are
observables to be found on which coherence is not small.
Since Bell was of the opinion that 
a `wave packet reduction', even on closed systems, 
`takes over from the Schr\"odinger equation', this was not to his
satisfaction.
He did agree however that these observables would become arbitrarily 
difficult to observe in practice for large $t$.

By considering large but finite closed systems subject to unitary
time evolution, we hope to clarify the role that macroscopic systems 
play in making us mistake coherent superpositions for classical mixtures. 
It seems striking that the same, simple mathematics can also
be used to understand why open systems do undergo decoherence as soon as
they lose information.


\begin{thebibliography}{Wer}
\bibitem[Arn]{Arn} M. Arndt et al., \textsl{Wave--Particle Duality of 
$C_{60}$--Molecules}, Nature {\bf 401}, 680--682, (1999).
\bibitem[Bel]{Bell} J. Bell, \textsl{`On Wave Packet Reduction in the 
Coleman--Hepp Model'}, Helv. Phys. Acta {\bf 48}, 93--98, (1975).
\bibitem[Fri]{Fri} J. Friedman et al., \textsl{Quantum
Superposition of Distinct Macroscopic States},
Nature {\bf 406}, 43--46, (2000).
\bibitem[Hep]{Hepp} K. Hepp, \textsl{`Quantum Theory of Measurement and 
Macroscopic Observables'}, Helv. Phys. Acta {\bf 45}, 237--248, (1972). 
\bibitem[Jan]{scriptie} B. Janssens, \textsl{`Quantum Measurement, a
Coherent Description'}, {\tt arxiv.org/abs/quant-ph/0503009}, (2004). 
\bibitem[J\&Z]{JandZ} E. Joos, H. Zeh, \textsl{`The Emergence of Classical Properties
Through Interaction with the Environment'}, Z. Phys. B {\bf 59}, 223--243,
(1985).
\bibitem[Neu]{Neu} J. von Neumann, \textsl{`Mathematische Grundlagen der
Quantenmechanik'}, Springer-Verlag, (1932).
\bibitem[Sch]{Sch} E. Schr\"odinger, \textsl{`Die 
Gegenwartige Situation in der Quantenmechanik'}, Naturwissenschaften
{\bf 23}, 807--812, 823--828, 844--849, (1935).
\bibitem[Tak]{Takesaki} M. Takesaki, \textsl{Theory of Operator Algebras
I}, Springer-Verlag, (1979). 
\bibitem[Wal]{Wal} H. van der Wal et al., \textsl{Quantum
Superposition of Macroscopic Persistent--Current States},
Science {\bf 290}, 773--777, (2000).
\bibitem[Wer]{Wer} R. Werner, \textsl{`Quantum 
Information Theory -- an Invitation'}, Springer Tracts in Modern Physics
{\bf 173}, 14--57, (2001).
\bibitem[Zur]{Zur} W. Zurek, \textsl{`Environment-induced Superselection
Rules'}, Phys. Rev. D {\bf 26}, 1862--1880, (1982).
\end{thebibliography}
\end{document}